\documentclass{svjour3}                     
\usepackage{graphicx}
\usepackage{amsmath}
\usepackage{amssymb, amsfonts}
\begin{document}
\title{Hamiltonian of mean force and dissipative scalar field theory}

\author{Marjan Jafari  \and
       Fardin Kheirandish 
       }

\institute{M. Jafari \at
              Department of Physics, Faculty of Science, Imam Khomeini International University, P.O.Box 34148 - 96818, Ghazvin, Iran \\
              \email{m.jafari@sci.ikiu.ac.ir}           
           \and
           F. Kheirandish \at
              Department of Physics, University of Kurdistan, P.O.Box 66177-15175, Sanandaj, Iran \\
              \email{fkheirandish@yahoo.com}
}

\maketitle
\begin{abstract}
In the framework of the Hamiltonian of mean force, internal energy, free energy and entropy of a dissipative 
scalar field are obtained. 
\keywords{Hamiltonian of Mean force \and Dissipation \and Scalar field \and Fano technique \and Coherent states}
\PACS{05.40.Jc \and 05.40.-a \and 05.30.-d}
\end{abstract}
\section{Introduction}\label{Introduction}
\noindent
Dissipative scalar field theories appear in important problems in classical and quantum physics. For example in Casimir Physics there are some special geometries where electromagnetic field can be basically considered as two independent scalar fields. Then these scalar fields should be quantized in the presence of a nonhomogeneous magnetodielectric medium where absorption and dispersion properties are taken into account \cite{1-11,4-11,5-11,11}. For the classical case one can model the fluctuating media by a collection of scalar fields and study the fluctuation-induced forces among immersed objects in such a medium \cite{Ramin}. 

The main obstacle in quantizing a dissipative field is that a Lagrangian or a Hamiltonian is required that generate the quantum evolution of the system and this leads to difficulties in implementing the canonical commutation relations if we are not allowed to include a heath bath \cite{1-3,2-3}. Various approaches to quantize a dissipative system have been introduced. Among these approaches, the system plus reservoir scheme allows the quantization of the total system described by a time-independent Hamiltonian in a rigorous way.

Caldeira and Leggett proposed a prototype model of a heat bath or reservoir consisting of a collection of non-interacting harmonic oscillators with different mass and frequencies coupled to the main system \cite{12-kh,13-kh,7-11}. Starting from a Lagrangian based on the Caldeira-Leggett model one can study the quantum dynamics of a dissipative system by different methods like path integral techniques \cite{6-2,7-2}, canonical quantization \cite{10-2} and phenomenological methods \cite{9-2}. The choice of a particular method to quantize an open quantum system depends on the physical properties of the environment of the main system. These physical properties are are considered as boundary conditions imposed on the dynamical variables describing the system.

The main purpose of the present article is to investigate the thermal equilibrium properties of a dissipative scalar field in the framework of mean force Hamiltonian. In this scheme, the thermodynamic equilibrium free energy of the main system is defined as the difference between the free energy of the whole system and free energy of just environment \cite{18-kh}. The Hamiltonian of mean force is in fact an effective Hamiltonian to describe the equilibrium properties of the main system \cite{19-kh}. Here we first quantize a scalar field in the presence of a medium or reservoir linearly coupled to it by modeling the environment as a continuum of non-interacting harmonic oscillators. Then the total system is quantized in the framework of canonical quantization and the total Hamiltonian is diagonalized using Fano technique \cite{20-kh}. In the following the memory function or susceptibility function is defined in terms of the reservoir Green's function and coupling function obeying Kramers-Kronig relations \cite{21-kh}. By making use of the Hamiltonian of the mean force, the thermal-energy, free energy and entropy of the scalar field are calculated. Finally, having the Hamiltonian diagonalized, coherent states of the total system are defined and correlation functions and expectation values of the relevant dynamical variables are obtained.
\section{The total Lagrangian }\label{Classical dynamics}
\noindent
The Lagrangian density of a real massive scalar field $\phi(x,t)$ in $3+1$-dimensional space-time, interacting with an environment modeled by a real scalar field $Y_\omega(x,t)$, is
\begin{equation}
L(t)=L_s+L_m+L_{int}.
\end{equation}
The first term $L_s$ is the Lagrangian density of the main system
\begin{equation}
L_s  = \frac{1}{2} \partial_\mu\phi(x,t)\partial^\mu\phi(x,t)-\frac{1}{2} m^2 \,\phi^2(x,t),
\end{equation}
the second term, is the Lagrangian density of the reservoir or heat bath consisting of a continuum of harmonic oscillators known as  Hopfield model \cite{26-8},
\begin{equation}
L _m= \frac{1}{2}\int\limits_0^\infty d\omega\,[\dot{Y}_\omega ^2 (x,t) - \omega ^2 \, Y_\omega ^2 (x,t)],
\end{equation}
and the last term is the interaction between the scalar field and its reservoir
\begin{equation}
L_{{\mathop{\rm int}} }  = \int\limits_0^\infty  {d\omega }  f(\omega )\,\phi (x,t)\,Y_\omega (x,t),
\end{equation}
where the scalar field is coupled to the reservoir linearly and $f(\omega)$ is the coupling function between them. For simplicity we work in the reciprocal space and write the fields in terms of their spatial Fourier transforms. The range of the variable $k$ in the reciprocal space is restricted to the half space \cite{19-ch}, thus in the reciprocal half space the Lagrangian of the system can be written as
\begin{equation}
\underline{L}_s=\int\limits_0^\infty d^3k (\left| \dot{\phi}(k,t) \right|^2-\omega_k^2\left| \phi(k,t) \right|^2),
\end{equation}
where $\omega_k^2=(m^2+k^2)$.
\begin{equation}
\underline{L}_R=\int\limits_0^\infty d^3k (\left| \dot{Y}_\omega(k,t) \right|^2-\omega^2\left| Y_\omega(k,t) \right|^2).
\end{equation}
\begin{equation}
\underline{L}_{\mbox{int}}=\frac{1}{2}\int\limits_0^\infty d^k \int \limits_0^\infty d\omega\, f(\omega)(\phi(k,t)Y_\omega^*(k,t)+\phi^*(k,t)Y_\omega(k,t))
\end{equation}
and $X_\omega^*(k,t)=X_\omega(-k,t)$. We can obtain the classical equations of motion simply from Euler-Lagrange equations. For $X_\omega(k, t), \phi(k, t)$ we find
\begin{eqnarray}\label{5}
(\partial_t^2+\omega_k^2)\,\phi(k,t)={\int\limits_0^\infty {d\omega}}\,f(\omega)\,X_\omega(k,t), \\ \label{6}
(\partial_t^2+\omega^2)X_\omega(k,t)=f(\omega)\,\phi(k,t).
\end{eqnarray}
The formal solution of the equation (\ref{6}) is
\begin{equation}\label{8}
Y_\omega (k,t) = {\dot{Y}} _\omega (k,0)\frac{{\sin \omega t}}{\omega } + Y_\omega (k,0)\cos \omega t + \int\limits_0^t {dt'}\,\frac{{\sin \omega (t - t')}}{\omega }f(\omega )\,\phi(k,t'),
\end{equation}
the first term is the solution of the homogeneous equation and after quantization plays the role of a quantum noise operator, the second term is the particular solution given by the Green's function of the scalar field
\begin{equation}\label{Green}
 G_\omega(t-t')=\Theta (t-t')\,\frac{{\sin \omega (t - t')}}{\omega },
\end{equation}
where $\Theta(t)$ is the Heaviside step function. The dimensionless memory or response function $\chi(t)$ is defined by
\begin{equation}\label{i0}
\chi_k(t) =\frac{1}{\omega_k^2}\,\int\limits_0^\infty d\omega\,\frac{{\sin \omega t}}{\omega }\, f^2(\omega ).
\end{equation}
The real and imaginary parts of Fourier transform of response function $\tilde{\chi }_k(\omega)$ satisfy Kramers-Kronig relations. From definition (\ref{i0}) and making use of inverse sine transform, one finds
\begin{equation}\label{CC}
f(\omega ) = \sqrt {\frac{2\omega\omega_k^2}{\pi}\mbox{Im}[\tilde{\chi} _k(\omega )]}.
\end{equation}
So for a given susceptibility function, one can adjust coupling function according to (\ref{CC}). From (\ref{i0}), we find
\begin{equation}
 \tilde{\chi}_k(\omega ) = \frac{1}{\omega_k^2}\{\mbox{P} \int\limits_0^\infty  {d\xi } \frac{{f ^2(\xi)}}{{\xi ^2  - \omega ^2 }} + i\pi{\frac{f(\omega )}{2\omega }}\},
\end{equation}
where $\mbox{P}$ means principal value. Now by substituting (\ref{8}) into (\ref{5}), a classical Langevin equation is obtained for the main oscillator
\begin{equation}\label{LEq}
\ddot{\phi} (k,t) + \omega _k^2 \,\phi (k,t) -\omega_k^2\,\int\limits_0^t dt'\, \chi _{k} (t - t')\,\phi(k,t') = \zeta^N (k,t),
\end{equation}
where
\begin{equation}\label{CNF}
  \zeta^N (k,t) = \int_0^\infty d\omega\, f(\omega )\,\left(Y_{\omega} (k,0)\,\cos(\omega t)+\dot{Y}_{\omega} (k,0)\,\frac{\sin(\omega t)}{\omega}\right),
\end{equation}
is the classical noise force.
\section{Quantization and dynamics}\label{Quantum dynamics}
\noindent
To quantize the total system canonically we need the conjugate momenta to impose equal-time quantization rules on them. From Lagrangian defined in the previous section we find the conjugate momenta in half $k$-space as
\begin{eqnarray}
 \Pi _\omega (k,t) &=& \frac{\delta L}{\delta \dot{Y}^*_{\omega}(k,t)} = \dot{Y} _{\omega}(k,t), \nonumber\\
 \pi (k,t) &=& \frac{\partial L}{\partial \dot{\phi}(k,t) } = \dot{\phi} (k,t).
\end{eqnarray}
The equal-time quantization rules are
\begin{eqnarray}
&& \left[ {\hat \phi^{\dag} (k,t),\hat \pi (k',t)} \right] = i\hbar \,\delta(k-k'), \\
&& \left[ \hat Y_{\omega}^\dag (k,t),\hat \Pi_{\omega'} (k',t) \right] = i\hbar\, \delta(\omega-\omega')\,\delta (k-k'), \\
\end{eqnarray}
and all other equal-time commutation relations are zero. Having the Lagrangian and conjugate momenta we can find the corresponding Hamiltonian as
\begin{eqnarray}\label{100}
\underline{\mathcal{H}} &=&
\left[\left|\hat{\pi}(k,t)\right|^2+\omega_k^2\left|\hat{\phi}(k,t)\right|^2\right]+{\int\limits_0^\infty}{d\omega}\,(\left|\hat{\Pi}_\omega(k,t)\right|^2 +\omega^2\left|\hat{Y}_\omega(k,t)\right|^2)\nonumber\\
&-& {\int\limits_0^\infty}{d\omega}f(\omega)\,(\hat{\phi}^\dag(k,t)\,\hat{Y}_\omega(k,t)+\hat{\phi}(k,t)\,\hat{Y}_\omega^\dag(k,t)).
\end{eqnarray}
In Heisenberg picture, one finds the equations of motion as operator analogs of classical equations of motion (\ref{5}, \ref{6}). To have a better understanding of the quantum dynamics of the system and also the subsystems, we diagonalize the Hamiltonian using Fano diagonalization technique. For this purpose, we assume
\begin{equation}\label{58}
\hat{H} =\int\limits_0^\infty dk \int\limits_0^\infty d\omega\, \hbar\omega\, \hat{C}^{\dag} (k,\omega, t)\hat{C} (k,\omega, t),
\end{equation}
that is the total Hamiltonian is assumed to be a continuum of uncoupled harmonic oscillators with creation and annihilation operators $\hat{C}^{\dag}$ and $\hat{C}$, respectively. These ladder operators satisfy bosonic commutation relations
\begin{equation}\label{66}
\left[ \hat{C}(k,\omega ,t),\hat{C}^\dag (k',\omega',t) \right] = \delta (\omega-\omega')\,\delta(k,k'),\,\,\,\,\,\,\left[ \hat{C}(k,\omega ,t),\hat{C}(k',\omega ',t)\right] = 0.
\end{equation}
The time-evolution of the annihilation and creation operators can be determined from Heisenberg equations using Hamiltonian (\ref{58})
\begin{eqnarray}
\hat{C}(k,\omega,t)=\hat{C}(k,\omega)e^{-i\omega t},\nonumber\\
\hat{C}^\dag(k,\omega,t)=\hat{C}^\dag(k,\omega)e^{i\omega t}.
\end{eqnarray}
Generally, the annihilation operator $\hat{C}$, can be considered a linear combination of the original dynamical degrees of freedom
\begin{eqnarray}
\hat{C} (k,\omega,t) = &-& \frac{i}{\hbar }\bigg[f_1^* (k,\omega )\hat{\phi}(k,t) -
f_2^* (k,\omega )\hat{\pi}(k,t)\nonumber \\
&+& \int\limits_0^\infty d\omega'\,f_3^*(k,\omega ,\omega')\hat{Y}_{\omega'} (k,t)-
\int\limits_0^\infty d\omega'\, f^*_4 (k,\omega,\omega')\hat{\Pi}_{\omega'} (k,t)\bigg]. \nonumber\\
\end{eqnarray}
To determine the coefficients $f_i$, Fano's diagonalization procedure is applied \cite{20-kh}, where commutator $[\hat{C}(k,\omega),H]$ is evaluated and the result is equaled to $ \omega\,\hat{C}(k,\omega)$. Comparing the contributions involving the various canonical operators we will find
\begin{equation}\label{gg1}
({\omega'}^2  - \omega ^2 )\,f_4 (\omega,\omega',k) =f_2 (\omega,k )\,f(\omega '),
\end{equation}
\begin{equation}\label{gg2}
(\omega_k^2  - \omega ^2 )\,f_2 (\omega,k) = \int\limits_0^\infty  d\omega'\, f_4 (\omega,\omega',k)\,f(\omega').
\end{equation}
Equations (\ref{gg1}, \ref{gg2}) are similar to Eqs.(\ref{5}, \ref{6}) in frequency domain, so using (\ref{8}), the general solution of these equations can be written as
\begin{eqnarray}\label{73}
f_4 (\omega ,\omega ',k) &=& h_4 (\omega,k)\,\delta (\omega  - \omega ') \nonumber\\
&&+ {\frac{{f(\omega ')}}{{2\omega '}}} \left( {\frac{{ 1}}{{\omega ' - \omega  - i0^ +  }} + \frac{{1}}{{\omega ' + \omega }}} \right)f_2 (\omega,k),
\end{eqnarray}
where $h_{4}(\omega,k)$ is an arbitrary function. The general solution to $f_{2}(\omega,k)$ is
\begin{equation}
f_2 (\omega ,k) = h_2 (\omega ,k) +  f(\omega )h_4(\omega,k )G_k (\omega ),
\end{equation}
where the Green's function $G$ is defined by the inverse of the $\Lambda, (G=\Lambda^{-1})$
\begin{equation}\label{75}
\Lambda_k (\omega ) = \left[(\omega_k ^2  - \omega^2 ) - P\int\limits_0^\infty d\xi \,\frac{f^2 (\omega )}
{\xi ^2  - \omega ^2 } - i\pi\,\frac{f^2(\omega )}{2\omega}\right],
\end{equation}
\begin{equation}\label{Gkapa}
G_k (\omega ) = \frac{-1}{\omega^2 -\omega _k^2 \,[1-\tilde{\chi} (\omega )]},
\end{equation}
and $h_{2}$ is the solution of
\begin{eqnarray}\label{hq}
\left[(\omega_k ^2  - \omega^2 )- P\int\limits_0^\infty d\xi \,\frac{f^2 (\omega )}
{\xi ^2  - \omega ^2 } - i\pi\,\frac{f^2(\omega )}{2\omega}\right] h_2 (\omega ) = 0.\nonumber\\
\end{eqnarray}
The explicit form of the function $h_4(\omega)$ can be determined from the commutation relations (\ref{66}) and (\ref{73}) as
\begin{equation}\label{79}
h_4 (\omega ) = \left( {\frac{\hbar }{{2\omega }}} \right)^{\frac{1}{2}}.
\end{equation}
The consistency of the diagonalization process leads to the following choice of $h_2(\omega)$
\begin{equation}\label{82}
h_2(\omega)=0.
\end{equation}
The set of coefficients of the diagonalizing transformation is now determined by Eqs. (\ref{73}), (\ref{75}), (\ref{79}) and (\ref{82}). The canonical operators in terms of the annihilation and creation operators can also be written in frequency domain. We have
\begin{equation}
\hat{\phi}(k,t)=\frac{1}{2\pi}\int\limits_0^\infty d\omega\,[\hat{\phi}(k,\omega)\exp(-i\omega t)+h.c.],
\end{equation}
\begin{equation}\label{phi}
\hat{\phi} (k,\omega) = 2\pi \left( \frac{\hbar }{2\omega } \right)^{\frac{1}{2}} f(\omega )G_k (\omega ) \hat{C} (\omega,k ) = \frac{i}{\omega }\hat{\pi} (k,\omega),
\end{equation}
and
\begin{eqnarray}\label{nnn}
\hat{Y}_{\omega} (k, \omega') &=& 2\pi \sqrt{\frac{\hbar}{2\omega}} {\delta (\omega  - \omega ')\,\hat C (k,\omega )}\nonumber\\
 && +  {\frac{{f(\omega )}}{{2\omega }}\left(\frac{1}{\omega  - \omega ' - i0^ +  } + \frac{1}{\omega  + \omega'}\right)\,\hat{\phi} (k,\omega')}  = \frac{i}{{\omega '}}\,\hat \Pi _{\omega} (k,\omega').\nonumber\\
\end{eqnarray}
\section{Thermal equilibrium}
\noindent
Having diagonalized Hamiltonian, now we proceed and find the thermal equilibrium expectation values of the internal energy and free energy of the main system in the framework of Hamiltonian of mean force. In global thermal equilibrium, we have
\begin{eqnarray}\label{222}
&& \langle \hat C^{\dag} (k,\omega )\,\hat C (k',\omega')\rangle = N(\omega )\delta (\omega-\omega')\,\delta(k-k')\\
&& \langle \hat C(k,\omega )\,\hat C(k',\omega')\rangle =0.
\end{eqnarray}
where $ N(\omega)=\exp (\hbar\omega/K_B T)-1$. Using Eq.(\ref{222}) and straightforward calculations, the symmetric thermal correlation functions of $\phi(r,t)$ are found as
\begin{eqnarray}
\langle\hat{\phi}(r,t)\hat{\phi}(r',t') \rangle &=&\frac{1}{(4\pi)^4}\int\limits_0^\infty d\omega\,\int\limits_0^\infty d\omega' \, \int\limits_0^\infty dk \,\int\limits_0^\infty dk' \, \bigg[ \exp[-i(\omega t-kr-\omega' t'+k'r')]\nonumber\\
&\times& \langle\hat{\phi}(k,\omega)\hat{\phi}^\dag(k',\omega') \rangle
+\exp[i(\omega t-kr-\omega' t'+k'r')] \langle\hat{\phi}^\dag(k,\omega)\hat{\phi}(k',\omega') \rangle \bigg].\nonumber\\
\end{eqnarray}
The frequency-domain correlation functions $\langle\hat{\phi}^\dag(k,\omega)\hat{\phi}(k',\omega') \rangle $ and $\langle\hat{\phi}(k,\omega)\hat{\phi}^\dag(k',\omega') \rangle $ are
\begin{eqnarray}
\langle\hat{\phi}^\dag(k,\omega)\hat{\phi}(k',\omega') \rangle &=&\frac{2\pi^2 \hbar}{\omega}f^2(\omega) G^*_k(\omega)G_{k'}(\omega')N(\omega)\delta(\omega-\omega')\delta(k-k'),\nonumber\\
&=&\frac{N(\omega)}{N(\omega)+1} \langle\hat{\phi}(k,\omega)\hat{\phi}^\dag(k',\omega') \rangle,
\end{eqnarray}
so for the symmetric field correlation function we find
\begin{eqnarray}
\frac{1}{2}\langle\hat{\phi}^\dag(r,t)\hat{\phi}(r',t')+\hat{\phi}(r,t)\hat{\phi}^\dag(r',t') \rangle= \nonumber\\
\frac{\hbar}{4\pi^3} \int\limits_0^\infty d\omega\,\int \limits_0^\infty dk
\coth(\frac{\hbar \omega}{2K_B T})\,\cos[\omega(t-t')-k(r-r')] \rm{Im}G_k(\omega),
\end{eqnarray}
and similarly for symmetric momentum correlation function
\begin{eqnarray}
\frac{1}{2}\langle\hat{\pi}^\dag(r,t)\hat{\pi}(r',t')+\hat{\pi}(r,t)\hat{\pi}^\dag(r',t') \rangle= \nonumber\\
\frac{\hbar}{4\pi^3} \int\limits_0^\infty d\omega\,\int \limits_0^\infty dk \,
\omega^2\,\coth(\frac{\hbar \omega}{2K_B T})\,\cos[\omega(t-t')-k(r-r')] \mbox{Im}G_k(\omega).
\end{eqnarray}
Having the explicit forms of the fields we can now find the thermal expectation values of the reservoir and interaction term as
\begin{eqnarray}
 && \frac{1}{2}\int\limits_0^\infty d\omega \,\langle (\partial_t \hat Y_\omega (r,t))^2  + \omega^2 \,\hat Y_\omega^2(r,t) \rangle\nonumber\\
 && = \frac{\hbar}{2\pi}\mbox{Im} \int\limits_0^\infty d\omega\,\int\limits_0^\infty dk\,\omega_k^2 \coth \left(\frac{\hbar \omega}{2K_B T}\right)\,
 \frac{d[\omega \,\chi_k(\omega )]}{d\omega}\, G_k (\omega ),
\end{eqnarray}
and
\begin{eqnarray}
&& \int\limits_0^\infty d\omega\, f(\omega )\langle \hat{\phi}(r,t)\,\hat{Y}_{\omega} (r,t) \rangle\nonumber\\
&& = \frac{\hbar}{2\pi}\mbox{Im} \int\limits_0^\infty d\omega \int\limits_0^\infty dk\,\omega_k^2\, \coth \left(\frac{\hbar \omega}{2K_B T} \right)\chi_k(\omega )\,G_k (\omega ).
\end{eqnarray}
\subsection{ Hamiltonian of mean force} \label{ Hamiltonian of mean force}
\noindent
Consider the total Hamiltonian of the system plus reservoir as
\begin{equation}
 \hat H=\hat H_S+\hat H_R +\hat H_I,
\end{equation}
where $\hat H_I$ is the interaction term and $H_s$ are $H_R$ are Hamiltonian of system and reservoir respectively. The density operator of the total system is defined by
\begin{equation}\label{density}
  \rho=\frac{e^{-\beta H}}{Z},
\end{equation}
where $Z=\mbox{tr} \{\exp (-\beta H)\}$ is the the total partition function. The reduced density operator is given by $\rho_s=\mbox{tr}_R \{\rho\}$ and when the interaction term is negligible the reduced density operator is given by
\begin{equation}\label{reduced}
  \rho_s=\frac{e^{-\beta H_S}}{Z_S},
\end{equation}
with $Z_s=\mbox{tr}_S \{\exp (-\beta H_S)\}$. When the coupling between the system and its environment is not negligible then the reduced density matrix can not be written as (\ref{reduced}) but generically can be written as
\begin{equation}\label{MFD}
  \rho_s=\frac{e^{-\beta H^*_S}}{Z^*_S},
\end{equation}
where the Hamiltonian of mean force or effective Hamiltonian is given by
\begin{equation}
\hat H_S^* =- \frac{1}{\beta}\,\ln \left( \frac{\mbox{tr}_R [e^{-\beta \hat H}]}{Z_R} \right),
 \end{equation}
and $\beta=1/K_B T$. The function $Z_R  = \mbox{tr}_R [\exp (-\beta \hat H_R )]$ is the partition function of the reservoir. The partition function $Z^*$ associated with the Hamiltonian of mean force $\hat H^*$ is defined by
\begin{equation}\label{part}
Z^* = \mbox{tr}_S [\exp (-\beta \hat H_S^* )] = \frac{Z}{{Z_R }}.
\end{equation}
From (\ref{part}) we define the free energy of mean force as
\begin{equation}
F^*  =  -\frac{1}{\beta}\,\ln (Z^* )= U - U_R - T(S - S_R ),
\end{equation}
where $U$ and $S$ are internal energy and entropy of total system. Now we can define the internal energy of mean force as
\begin{equation}\label{internal}
U^* = U- U_R  = \langle {\hat H}\rangle_{\mbox{tot}} - Z_R^{-1} \mbox{tr}_R [\hat H_R e^{-\beta \hat H_R }].
\end{equation}
\subsection{Free energy and entropy}
\noindent
The free energy of mean force can be obtained from $U^*=-T^2 \,\partial_T (F^* /T)$, as
\begin{eqnarray}
F^* &=& \frac{K_B T}{\pi}\int\limits_0^\infty d\omega\,\int\limits_0^\infty dk\, \ln \left[\sinh \left(\frac{\hbar \omega}{2K_B T} \right) \right]\,\mbox{Im}
\left\{\frac{\omega_k^2 \left[\omega\frac{d\,\chi_k(\omega)}{d\omega}- \chi_k(\omega)+ 1 \right] + \omega^2 }{\omega_k^2
[1-\chi_k(\omega)]-\omega^2}\right\}\nonumber\\
&+& K_B T\,\ln 2,
\end{eqnarray}
now using the standard thermodynamic relation, $S^*=-\partial_T F^*$, the entropy of mean force is obtained as
\begin{eqnarray}
S^* &=& \frac{K_B T}{\pi}\int\limits_0^\infty d\omega\,\int\limits_0^\infty dk\, \left\{\frac{1}{T}\,\coth\left(\frac{\hbar \omega}{2 K_B T}\right)-\frac{2 K_B}{\hbar \omega}\,
\ln \left[\sinh\left(\frac{\hbar \omega}{2 K_B T}\right)\right] \right\}\nonumber\\
&\times & \left[\mbox{Im}
\left\{\frac{\omega_k^2 \left[\omega\frac{d\chi_k(\omega)}{d\omega}- \chi_k(\omega)+ 1 \right] + \omega^2 }{\omega_k^2
[1-\chi_k(\omega)]-\omega^2}\right\}\right]+ K_B T\,\ln 2.
\end{eqnarray}
\section{Coherent states}
\noindent
In contrast to a thermal state which is a statistical mixture of bosons, a coherent state $\left| C(\omega) \right\rangle $, also called a Glauber state \cite{coherent}, is the nearest state to classical behaviour. Coherent states can be defined as the eigenstates of the annihilation operator $\hat{C}(\omega)$
\begin{equation}
\hat{C(k,\omega)}\left| C(k,\omega) \right\rangle= C(k,\omega) \left| C(k,\omega) \right\rangle,
\end{equation}
which is the standard definition of coherent states, the reader can consult \cite{prelomov} for other equivalent definitions. Basic coherent state expectation values are
\begin{eqnarray}\label{chek}
\left\langle C(k,\omega) \right| \hat{C}^\dag(k,\omega)\hat{C}(k',\omega') \left| C(k,\omega)\right\rangle= \left| C(k,\omega)\right|^2\delta(k-k')\delta(\omega-\omega')\nonumber\\
=\left\langle C(k,\omega) \right| \hat{C}(k,\omega)\hat{C}^\dag(k',\omega') \left| C(k,\omega)\right\rangle-\delta(k-k')\delta(\omega-\omega')\\\label{ccoh}
\left\langle C(k,\omega) \right| \hat{C}(k,\omega)\hat{C}(k,\omega) \left| C(k,\omega)\right\rangle= C^2(k,\omega)\delta(k-k')\delta(\omega-\omega')\\
\left\langle C(k,\omega) \right| \hat{C}^\dag(k,\omega)\hat{C}^\dag(k,\omega) \left| C(k,\omega)\right\rangle= {C^*}^2(k,\omega)\delta(k-k')\delta(\omega-\omega')\label{coh}.
\end{eqnarray}
For the expectation value of the normal-mode $\hat{\phi}(k,\omega)$, and its canonical conjugate $\hat{\pi}(k,\omega)$ we find
\begin{eqnarray}
&& \left\langle C(k,\omega) \right| \hat{\phi}(k,\omega) \left| C(k,\omega)\right\rangle=2\pi \sqrt{\frac{\hbar}{2\omega}}(f(\omega)G_k(\omega)C(k,\omega)+f^*(\omega)G_k^*(\omega)C^*(k,\omega)),\nonumber\\
\\
&& \left\langle C(k,\omega) \right| \hat{\pi}(k,\omega) \left| C(k,\omega)\right\rangle=-i2\pi \sqrt{\frac{\hbar\omega}{2}}(f(\omega)G_k(\omega)C(k,\omega)-f^*(\omega)G_k^*(\omega)C^*(k,\omega)).\nonumber\\
\end{eqnarray}
It follows from (\ref{chek}), (\ref{ccoh}) and (\ref{coh}) that the coherent auto-correlation function of $\hat{\phi(r,t)}$ can be written as
\begin{eqnarray}
&&\left\langle \hat{\phi}(r,t)\hat{\phi}(r',t')\right\rangle_{coh}=\nonumber\\
&&\frac{1}{8\pi^2}\int\limits_0^\infty d\omega \int \limits _0^\infty dk \cos[k(r-r')-t(\omega-\omega')]\,(\left\langle \phi(k,\omega)\right\rangle_{coh}^2+4\pi \hbar\,Im G_k(\omega)),\nonumber\\
\end{eqnarray}
and for the conjugate field
\begin{eqnarray}
&&\left\langle \hat{\pi}(r,t)\hat{\pi}(r',t')\right\rangle_{coh}=\nonumber\\
&&\frac{1}{8\pi^2}\int\limits_0^\infty d\omega \int \limits _0^\infty dk \cos[k(r-r')-t(\omega-\omega')]\,(-\left\langle \pi(k,\omega)\right\rangle_{coh}^2+4\pi \hbar\,\omega^2\,Im G_k(\omega)).\nonumber\\
\end{eqnarray}
\section{Conclusion}
\noindent
Starting from a Lagrangian, quantum dynamics of a dissipative scalar field was investigated in the frame work of Huttner-Barnett model. The Hamiltonian was derived and diagonalized using Fano diagonalization technique. The memory function or susceptibility of the system was defined in terms of the reservoir Green's function ant the coupling function between the system and its environment. In the framework of the Hamiltonian of mean force and in thermal equilibrium, internal energy, free energy and entropy of the scalar field were obtained. Correlation functions in thermal and coherent states were found.
\section*{References}

\end{document}